
\input phyzzx.tex
\date{CLNS 91/1118}
\titlepage
\title{\bf One-Point Functions of Loops and
Constraints Equations of the Multi-Matrix Models at finite
$N$}
\author{Changrim Ahn
\foot{E-mail address: ahn@cornella.bitnet}}
\address{F. R. Newman Lab. of Nuclear Studies,\break
                  Cornell University \break Ithaca,NY 14853}
\andauthor{Kazuyasu Shigemoto
\foot{E-mail address: shigemot@jpnrifp.bitnet}}
\address{Department of Physics,\break Tezukayama University
                                  \break Nara 631, Japan}

\abstract{
We derive one-point functions of the loop operators of
Hermitian matrix-chain models at finite $N$ in terms of
differential operators acting on the partition functions.
The differential operators are completely determined by recursion
relations from the Schwinger-Dyson equations. Interesting observation is
that these
generating operators of the one-point functions
satisfy
$W_{1+\infty}$-like algebra.
Also, we obtain constraint equations on the partition
functions in terms of the differential operators. These constraint
equations on the partition functions define the symmetries of the matrix
models at off-critical point before taking the double scaling limit.
}
\endpage
\def\tr{{\rm Tr}}
\def\V{{\cal V}}
\def\pd{\partial}

\def\[{\Bigl\langle}
\def\]{\Bigr\rangle}
\def\({\Big(}
\def\){\Big)}
\def\:{{\scriptscriptstyle ^{^{^\times}}_{{\times}}}}
\def\cd{{\scriptscriptstyle ^\circ}}
\def\W{{\cal W}}
\REF\Brezin{E. Brezin and V. Kazakov, Phys. Lett. {\bf B236} (1990)
144;
M. Douglas and S. Shenker, Nucl. Phys. {\bf B335} (1990)
635;
D. Gross and A. Migdal, Phys. Rev. Lett. {\bf 64} (1990)
127.}
\REF\Rutgers{T. Banks, M. Douglas, N. Seiberg, and S. Shenker, Phys.
Lett.
{\bf B238} (1990) 279.}
\REF\Douglas{M. Douglas, Phys. Lett. {\bf B238} (1990) 176.}
\REF\Lebedev{A. Gerasimov, A.Marshakov, A. Mironov and A. Orlov, Nucl.
Phys.
{\bf B357} (1991) 565.}
\REF\CERN{L. Alvarez-Gaume, C. Gomez and J. Lacki, Phys. Lett. {\bf
B253}
(1991) 56.}
\REF\Martinec{E. Martinec, Commun. Math. Phys. {\bf 138} (1991) 437.}
\REF\Ours{C. Ahn and K. Shigemoto, Phys. Lett. {\bf B263} (1991) 44.}
\REF\Kawai{M. Fukuma, H. Kawai, and R. Nakayama,
Int. J. Mod. Phys. {\bf A6} (1991) 1385;
R. Dijkgraaf, E. Verlinde, and H. Verlinde, Nucl. Phys.
{\bf B348} (1991) 435.}
\REF\Goeree{J. Goeree, Nucl. Phys. {\bf B358} (1991) 737;
K. Li, Nucl. Phys. {\bf B354} (1991) 725.}
\REF\Gava{E. Gava and K.S. Narain, Phys. Lett. {\bf B263} (1991) 213.}
\REF\Mironov{A. Mironov and A. Morozov, Phys. Lett. {\bf B252} (1990)
47.}
\REF\Itoyama{H. Itoyama and Y. Matsuo, Phys. Lett. {\bf B255} (1991)
202.}
\REF\Bakas{I. Bakas, Phys. Lett. {\bf B228} (1989) 57.}
\REF\Tada{T. Tada, Phys. Lett. {\bf B259} (1991) 442;
M. Douglas, In 1990 Cargese conference.}
\REF\Itoyamaii{H. Itoyama and Y. Matsuo, Phys. Lett. {\bf B262} (1991)
233.}
\REF\Sin{M. A. Awada and S. J. Sin, U. of Florida Preprint 90-33,
91-3.}
\REF\Cheng{Y.-X. Cheng, Osaka U. Preprint OU-HET 159.}
\noindent
{\bf 1. Introduction}
\hfill

Recently much progress has been made on the matrix model formulation of
the
2D gravity to study the non-perturbative
effects,\refmark{\Brezin}\
and an interesting connection with the
integrable systems has been made in the double scaling
limit, in which the size of the matrix $N$ becomes infinite and the
matrix
potentials have critical forms.\refmark{\Brezin}\
In this limit, the non-perturbative results can be obtained from
non-linear integrable differential equations, such as KdV equations.
Furthermore, the correlation functions satisfy their hierarchical
equations.\refmark{\Rutgers-\Douglas}\

It has been noticed recently that the integrability of the matrix models
is maintained even at off-critical points (finite $N$) before taking the
double scaling limit.
At finite $N$, the Lax pair, zero-curvature conditions, and infinite
number of conserved quantities of the matrix model have been derived
and related to integrable systems in more clear and direct way.
The underlying integrable systems have been
identified with 1D Toda hierarchy for one-matrix model\refmark\Lebedev\
which becomes the
KdV hierarchy in the scaling limit,\refmark\CERN\
2D Toda hierarchy,\refmark{\Lebedev,\Martinec}\ and 2D Toda
multi-component
hierarchy\refmark\Ours\ for the two-matrix model and
for the general multi-matrix models, respectively.
The partition functions are the `$\tau$-functions' of these integrable
systems.

Next object of interest is the correlation functions of
local operators. For the operators appearing in the action, the
correlation functions are simply given by the derivatives of the
partition functions with respect to the coefficients of the operators in
the action. It requires, however, non-trivial analysis for the operators
which do not appear in the action.
In this paper, we derive one-point functions for the general local
operators which are the `loops' in the matrix models in terms of
differential operators acting on the partition functions. These
operators satisfy the recursion relations coming from the
Schwinger-Dyson equations.
We notice that the commutation relations of these differential operators
are similar to those of the $W_{1+\infty}$ algebra and become exact in
the continuum limit.

One important related problem is the symmetry structure of the matrix
models.
The Virasoro and $W_{p+1}$ algebras
have been conjectured for the $p$ multi-matrix models
as constraint equations on the partition functions
in the double scaling limit.\refmark{\Kawai-\Goeree}\
The derivation of these symmetries, however, has not been made except
for the one-matrix model (the Virasoro algebra)
and a special two-matrix model\refmark\Gava\ (the $W_3$ algebra).
This derivation may be possible if one consider the constraint equations
of the matrix models at finite $N$ first.
Indeed, it is at finite $N$ that the Virasoro algebras have been derived for
the one-matrix model\refmark{\CERN,\Mironov, \Itoyama}\ and
for the multi-matrix model.\refmark\Ours\
In this paper, we derive most general constraint equations for the
multi-matrix models in terms of the generators of the
$W_{1+\infty}$-like algebra.
These constraint
equations seem to be consistent with the conjectures made in the double
scaling limit.

\noindent
{\bf 2. Two-Matrix Model}\hfill

The partition function of the Hermian two-matrix model is given by
$$\eqalign{&Z\left[\{t_k\};\{s_k\},c\right]
=\int {\cal D}U {\cal D}V e^{-S},\cr
&S= \V_1(U) +\V_2(V)-cUV,\quad \V_1(U)=\sum_{k=1}^{\infty} t_k U^k,\quad
\V_2(V)=\sum_{k=1}^{q} s_k V^k.\cr}\eqn\PFi$$
Note the difference in the two potentials $\V_1$ and $\V_2$;
$\V_1$ is arbitrary polynomial potential and $\V_2$ is with fixed order.
We want to express correlation functions in terms of $t_k$'s and their
derivatives acting on the partition functions. These differential
operators depend explicitly on the another parameters, $s_k$'s.

The most interesting loop operators in the two-matrix models are
$\tr(V^{n} U^m)$.
The one-point functions of these loops are
$$\[\tr\(V^{n} U^m\)\]
=\int{\cal D}U {\cal D}V e^{-S}
\left[\tr\(V^{n} U^{m}\)\right].\eqn\MPFi$$
{}From the Schwinger-Dyson (SD) equations,
$$\sum_{i,j=1}^N \int{\cal D}U {\cal D}V {\partial\over{\partial
X_{ij}}}\left[\(V^n U^m\)_{ij} e^{-S}\right]=0,\quad (X=U,V),
\quad (m,n\ge 0),\eqn\SDEi$$
one can derive two recursion relations as follows:
$$\eqalignno{
&c\[\tr\(V^{n+1} U^m\)\]=\[\tr\(V^n U^m\V_1^{\prime}(U)\)\]
-\sum_{r=0}^{m-1}\[\tr\(V^n U^{m-r-1}\)\tr U^r\],&(4)\cr
&c\[\tr\(V^n U^{m+1}\)\]=\[\tr\(V^n\V_2^{\prime}(V) U^m\)\]
-\sum_{s=0}^{n-1}\[\tr V^s\tr\(V^{n-s-1} U^m\)\].&(5)\cr}$$

\equanumber=5

Differential operators generating the one-point functions, defined by
$$W^{(n+1)}_{m-n}Z\left[\{t_k\};\{s_k\},c\right]
\equiv -c^n\[\tr\(V^n U^m\)\],\eqn\Wgen$$
satisfy the recursion relation from Eq.(4),
$$\eqalign{W^{(n+2)}_{m-n-1}&=\sum_{r=0}^{m-1}
{\partial\over{\partial t_r}}\cd W^{(n+1)}_{m-n-r-1}
+ \sum_{r=1}^{\infty}r t_r W^{(n+1)}_{m-n+r-1},\cr
W^{(1)}_m&={\partial\over{\partial t_m}},\quad {\partial\over{\partial
t_0}}\equiv -N,\quad{\rm for}\quad m,n\ge 0,\cr}
\eqn\REQiii$$
where the symbol $\cd$ is defined by $(A\cd B)Z=A(BZ)$.
This recursion relation can be rewritten in a simple form
$$W^{(n+1)}_{m}=\sum_{r=-\infty}^{m+n-1}\: J_r\cd
W^{(n)}_{m-r}\:\ (n\ge 0,m\ge -n),\quad
J_r=\cases{\partial/\partial t_r,&if $r>0$\cr
rt_r,&if $r<0$\cr},\eqn\REQiv$$
where $\:\cdots\:$ denotes the normal ordering. The operator $J_r$'s satisfy
$U(1)$ current algebra
$[J_m,J_n]=m\delta_{m+n,0}$ ($J(z)=\sum_{m} J_m z^{-m+1}=\partial_z\phi $ ).
Eq.\REQiv\ defines recursively the generating differential operators
of the one-point functions. If there is no upper limit in the summation range
($m+n\to\infty$), it is obvious that
$$W^{(n)}(z)=\sum_{m} W^{(n)}_m z^{-m+n}=
\:\(\partial_z\phi\)^n\:.\eqn\xxx$$
These infinite number of currents $W^{(n)}(z)$ ($n=1,2,\cdots$) generate
the $W_{1+\infty}$ algebra.\refmark\Bakas\
For the finite values of $m+n$, however, the commution relations are
not exactly same as those of the $W_{1+\infty}$ algebra.
This $W_{1+\infty}$-like algebra generates the one-point functions of
the loops.
It is remarkable that in the continuum limit the loop operators
are given by the operators like $\tr(U^M)$ with `lattice spacing' $a\to 0$
and `lattice size'
$M\to\infty$ while keeping $aM$ finite. Therefore, the
$W_{1+\infty}$ algebra generates the one-point functions in the double
scaling limit.

For explicit examples and later use,
we write explicit expressions for $W^{(2)}_m,W^{(3)}_m$,
$$\eqalign{W_{m}^{(2)}
&=\sum_{r=0}^{m} \partial_r \partial_{m-r}+\sum_{r>0} r t_r
\partial_{m+r},\cr
W_{m}^{(3)}
&=\sum_{r=0}^{m+1}\sum_{s=0}^{m-r}
\partial_r \partial_s \partial_{m-r-s}
+\sum_{r>0} r t_r\left(\sum_{s=0}^{m+1}\partial_s \partial_{m+r-s}
+\sum_{s=0}^{m+r} \partial_s \partial_{m+r-s}\right)\cr
&+\sum_{r,s=0}^{\infty} r t_r s t_s \partial_{m+r+s}
+{{(m+2)(m+1)} \over 2} \partial_{m},\cr}\eqn\explicit$$
where $\partial_m=\partial/\partial t_m$ and $W_{m}^{(2)}$ can be
identified with the Virasoro generator $L_m$ as it satisfies the classical
Virasoro algebra $[L_m, L_n]=(m-n)L_{m+n}$.

Now consider the constraint equations on the partition
function. One can express Eq.(5) with the one-point generating operators
as follows:
$$\eqalign{&{\widehat W}_{m}^{(n+1)}Z[\{t_k\};\{s_k\},c]=0,
\quad{\rm for}\quad n\ge 0,\ m\ge -n,\cr
&{\widehat W}_{m}^{(n+1)}=W^{(n+1)}_m
-\sum_{k=1}^{q}{ks_k\over{c^k}}W^{(k+n)}_{m-k}
-\sum_{k=0}^{n-1}
W^{(n-k)}_{m+k}\cd W^{(k+1)}_{-k}.\cr}\eqn\CEQi$$
Not all of these constraints are independent. In fact, we can prove
that only ${\widehat W}_{m}^{(1)}$'s are independent by showing the
following relation from Eqs.(4) and \CEQi
$${\widehat W}_{m}^{(n+1)}=\sum_{r=-\infty}^{m+n-1}\:J_r\cd {\widehat
W}_{m-r}^{(n)}\: \quad (n\ge 0,\ m\ge -n) .\eqn\eq$$
If ${\widehat W}_{m}^{(1)}Z=0$, ${\widehat
W}_{m}^{(n>1)}Z=0$ are automatically satisfied.
Therefore, the constraint equations for the two-matrix models become
$${\widehat W}_{m}^{(1)}Z=\left[ {\partial\over{\partial t_{m}}}
 -\sum_{k=1}^{q}{ks_k\over{c^k}}W^{(k)}_{m-k}
\right]Z=0.\eqn\CEQii$$
As one can see in Eq.\CEQii, the constraints are linear
combinations of the generators of the
$W_{1+\infty}$-like algebra with the coefficients
of the second potential $\V_2$.
In the continuum limit, the constraint equations are given by the
$W_{1+\infty}$ algebra. Furthermore, for a special potential $\V_2$,
the operators ${\widehat W}_{m}^{(1)}$'s may generate a subalgebra of the
$W_{1+\infty}$, say, the $W_n$
algebra.

One can realize the $W_{1+\infty}$-like algebra as a symmetry of the matrix
model in the context of quantum field theory.
In the ordinary quantum field theory,
the symmetry
can be found as infinitesimal changes of the quantum fields which leave
the action invariant. For the matrix model, this can be done by the
following generators $A_{m,n},B_{m,n}$ defined by
$$\eqalign{
A_{m,n}\left[e^{-S}\right]&=\sum_{i,j=1}^N {\partial\over{\partial
U_{ij}}}\left[\(U^m V^n\)_{i,j}e^{-S}\right],\cr
B_{m,n}\left[e^{-S}\right]&=\sum_{i,j=1}^N {\partial\over{\partial
V_{ij}}}\left[\(U^m V^n\)_{i,j}e^{-S}\right].\cr}\eqn\ABgen$$
The generators $A_{m,n}$ and $B_{m,n}$ satisfy the following
closed commutation relations:
$$\eqalign{ [A_{m,0}, B_{k,l}] & = k B_{m+k-1,l},\quad
[A_{m,n}, B_{0,l}] = -n A_{m,n+l-1},\cr
[A_{m,1}, B_{1,l}]&= B_{m,l+1} -A_{m+1,l}.\cr}\eqn\ABcommu$$
Note that $\[A_{m,n}[e^{-S}]\]=\[B_{m,n}[e^{-S}]\]=0$
from the SD equations \SDEi, which become the constraint equations as
shown above.
This realization makes it simple to prove the statement that only
the $n=0$ constraints are independent. This comes from the fact
that $\[B_{m,n}[e^{-S}]\]=0$ can
be obtained from $\[B_{m,0}[e^{-S}]\]=0$ by using
the commutation relations \ABcommu.
In fact, it is not difficult to see that $\left\{A_{0,0},A_{2,1},
B_{1,0},B_{0,2}\right\}$ are enough to generate the constraints.

\noindent
{\bf 3. Multi-Matrix Models}\hfill

The multi-matrix models with $p$ Hermitian matrix
variables $U_a$ have the partition function
$$Z[\{t_k\};\{s_{a,k}\},\{c_a\}]
=\int \prod_{a=1}^p {\cal D} U_a e^{-S},\quad
S=\tr\left\{ \sum_{a=1}^p \V_a(U_a)-\sum_{a=1}^{p-1}c_a U_a U_{a+1}
\right\}.\eqn\action$$
We will choose the matrix potentials
$$\V_1(U_1)=\sum_{k=1}^{\infty} t_k U_1^k,\quad
\V_a(U_a)=\sum_{k=1}^{q_a}s_{a,k} U_a^k,\eqn\potential$$
considering $t_k$'s as variables and $s_{a,k}$'s as fixed parameters.
The loop operators in the multi-matrix models are given by
$\tr\left[U_p^{n_p}\cdots U_2^{n_2}U_1^{n_1}\right]$.
Again, we want to express one-point functions of these operators in
terms of the linear differential operators.

We start with the SD equations:
$$\eqalign{
&\sum_{i,j=1}^{N}\int[{\cal D}U] {\pd\over{\pd (U_{1})_{ij}}}
\left[\left(U_1^{n_1}
U_p^{n_p}\cdots U_2^{n_2}\right)_{ij}e^{-S}\right] =0,\cr
&\sum_{i,j=1}^{N}\int[{\cal D}U] {\pd\over{\pd (U_a)_{ij}}}
\left[\left(U_{a-1}^{n_{a-1}}\cdots U_1^{n_1} U_{p}^{n_{p}}\cdots
 U_{a+1}^{n_{a+1}}
\right)_{ij}e^{-S}\right] =0,(2\le a\le p-1)\cr
&\sum_{i,j=1}^{N}\int[{\cal D}U] {\pd\over{\pd (U_p)_{ij}}}
\left[\left(U_{p-1}^{n_{p-1}}\cdots U_{1}^{n_{1}} U_{p}^{n_{p}}
\right)_{ij}e^{-S}\right] =0.\cr}\eqn\SDEii$$

Eq.\SDEii\ can be used to derive recursion relations for the
one-point functions of the loop operators:
$$\eqalignno{
&c_1\[\tr\(U_p^{n_p}\cdots U_2^{n_2+1}U_1^{n_1}\)\]&\cr
&=-\sum_{r=0}^{n_1-1}\[\tr U_1^r\tr\left(
U_p^{n_p}\cdots U_2^{n_2}U_1^{n_1-r-1}\right)\]
+\[\tr\left(U_p^{n_p}\cdots U_2^{n_2}U_1^{n_1}\V_1^{\prime}(U_1)\right)\],
&(19)\cr
&c_{a-1}\[\tr\(U_{p}^{n_{p}}\cdots U_{a+1}^{n_{a+1}}U_{a-1}^{n_{a-1}+1}\cdots
U_1^{n_1}\)\] + c_a \[\tr\(U_{p}^{n_{p}}\cdots
U_{a+1}^{n_{a+1}+1}U_{a-1}^{n_{a-1}}
\cdots U_1^{n_1}\)\]&\cr
&=\[\tr\(U_{p}^{n_{p}}\cdots U_{a+1}^{n_{a+1}}\V_a^{\prime}(U_a)
U_{a-1}^{n_{a-1}}\cdots U_1^{n_1}\)\],
\qquad 2\le a\le p-1 &(20)\cr
&c_{p-1}\[\tr\(U_{p}^{n_{p}} U_{p-1}^{n_{p-1}+1}\cdots U_1^{n_1}\)\]
= -\sum_{r=0}^{n_p-1}\[\tr U_p^r\tr\left(
U_p^{n_p-r-1}U_{p-1}^{n_{p-1}}\cdots U_1^{n_1}\right)\]& \cr
&+\[\tr\(\V_p^{\prime}(U_p) U_{p}^{n_{p}} U_{p-1}^{n_{p-1}} \cdots
U_1^{n_1}\)\].&(21)\cr}$$

\equanumber=21

Eq.(19) can be rewritten in the form of the recursion relations
in terms of the differential operators
$$\eqalign{&\W^{(n+1)}_{m}(n_3,\cdots,n_p)=\sum_{r=-\infty}^{m+n-1}\:J_r\cd
\W^{(n)}_{m-r}(n_3,\cdots,n_p)\:,\cr
&\W^{(n+1)}_{m-n}(n_3,\cdots,n_p)Z[\{t_k\}]
=-c_1^n c_2^{n_3}\cdots c_{p-1}^{n_p}
\[\tr\(U_p^{n_p}\cdots U_2^n U_1^{m}\)\].\cr}\eqn\MREQi$$
We want to show that any one-point function can be expressed in terms of
the variables $t_k$'s and their derivatives. Since Eq.\MREQi\ for
$n_3=\cdots=n_p=0$ with the identification $c=c_1$
is exactly same as Eq.\REQiv, $\W^{(n+1)}_{m}(0,\cdots,
0)= W^{(n+1)}_{m}$ of the two-matrix model.

To derive other one-point functions, we consider other recursion
formulae coming from Eq.(20), ($2\le a\le p-1$)
$$\eqalign{&\W^{(n+1)}_{m}(n_3,\cdots,n_{a-1},0,n_{a+1}+1,\cdots,n_p)\cr
=-&{c_{a-1} \over {c_{a-2}}}\W^{(n+1)}_{m}(n_3,\cdots,n_{a-1}+1,0,n_{a+1},
\cdots,n_p)\cr
+&\sum_{k=1}^{q_a} {ks_{a,k}\over{c_{a-1}^{k-1}}}
\W^{(n+1)}_{m}(n_3,\cdots,n_{a-1},k-1,n_{a+1},\cdots,n_p).\cr}\eqn\MREQii$$
Assuming that we can express all operators $\W^{(n)}_m(0,\cdots,0,n_a,n_{a+1},
\cdots,n_p)$ in terms of $t_k$'s, we can find
$\W^{(n)}_m(0,\cdots,0,n_{a+1},\cdots,n_p)$ by repeatedly using
Eq.\MREQii.
Therefore, we showed inductively that all the one-point functions can be
found as differential operators of $t_k$'s acting on the partition
function.
Finally, if one finds all the operators in terms of $t_k$'s, one can
find the constraint equations on the partition function from Eq.(21).
Among others, the case of $n_p=0$ gives the following equations:
$$\eqalign{&{\widehat\W}_m^{(1)}(n_3,\cdots,n_{p-1}+1,0)
Z[\{t_k\};\{s_{a,k}\},\{c_a\}]=0,\cr
&{\widehat\W}_m^{(1)}(n_3,\cdots,n_{p-1}+1,0)
=\W^{(1)}_{m}(n_3,\cdots,n_{p-1}+1,0) \cr
&-c_{p-2}\sum_{k=1}^{q_p}{ks_{p,k}\over{c_{p-1}^{k}}}
\W^{(1)}_{m}(n_3,\cdots,n_{p-1},k-1), \quad (p \ge 4)   \cr }
\eqn\CEQii$$
and we must treat carefully the index $n$ in Eqs.(23) and (24) for
$p \le 3$.

We apply above general analysis to the three-matrix model. Defining
$$\W^{(n+1)}_{m-n}(l)Z[\{t_k\}]
=-c_1^n c_2^l \[\tr\(W^l V^n U^m\)\],\eqn\eq$$
they satisfy the following recursion relations:
$$\eqalign{
&\W^{(n+1)}_{m}(l)=\sum_{r=-\infty}^{m+n-1}\:J_r\cd\W^{(n)}_{m-r}(l)\:,\cr
&\W^{(1)}_{m}(l+1)
=-c_1\W^{(1)}_{m+1}(l)
+\sum_{k=2}^{q_2} {ks_{2,k}\over{c_1^{k-1}}}
\W^{(k)}_{m}(l).\cr}\eqn\MREQiii$$
As explained above, from $\W^{(n)}_{m}(0)=W^{(n)}_{m}$ one can find
$\W^{(1)}_{m}(1)$'s from the second equation and $\W^{(n)}_{m}(l)$'s
from the first one. Continuing this step, one can find all
$\W^{(n)}_{m}(l)$'s. Finally, the constraint equations come from
Eq.\CEQii.
For an explicit example, consider the following potentials:
$\V_2(V)=v_2 V^2+ v_3 V^3$ and $\V_3(W)=w_2 W^2 + w_3 W^3$.
{}From Eq.\MREQiii, one can find the explicit expressions:
$$\eqalign{&\W^{(1)}_{m}(1)=- c_1 W^{(1)}_{m+1}+{2v_2\over{c_1}}W^{(2)}_{m}
+{3v_3\over{c_1^2}}W^{(3)}_{m}.\cr
&\W^{(2)}_{m}(1)=\sum_{r=0}^{m}
{\partial\over{\partial t_r}}\cd \W^{(1)}_{m-r}(1)
+ \sum_{r=1}^{\infty}r t_r \W^{(1)}_{m+r}(1),\cr
&\W^{(3)}_{m}(1)=\sum_{r=0}^{m+1}
{\partial\over{\partial t_r}}\cd \W^{(2)}_{m-r}(1)
 + \sum_{r=1}^{\infty}r t_r \W^{(2)}_{m+r}(1),\cr
&\W^{(1)}_{m}(2)=-c_1 \W^{(1)}_{m+1}(1) +{2v_2\over{c_1}}\W^{(2)}_{m}(1)
+{3v_3\over{c_1^2}}\W^{(3)}_{m}(1),\cr
}\eqn\eq$$
and the constraint equations are
$${\widehat{\cal W}}_{m-1}^{(2)}(0) Z
=\left[W^{(2)}_{m-1}-{2 c_1 w_2\over{c_2^2}}\W^{(1)}_{m}(1)
-{3 c_1 w_3\over{c_2^3}}\W^{(1)}_{m}(2)
\right]Z=0,\eqn\MCEQ$$
where $W^{(2)}_{m},W^{(3)}_{m}$ are given in Eq.\explicit.

\noindent
{\bf 4. Discussions}\hfill

In this paper, we computed one-point functions of the multi-matrix
models in terms of the differential operators acting on the partition
functions. The operators are completely determined by the recursion
formulae, derived from the SD equations and generate
the $W_{1+\infty}$-like algebra. Furthermore, we derived the
constraint equations on the partition functions using these operators.
Since the partition functions are the `$\tau$ functions' of the 2D Toda
hierarchies,\refmark{\Lebedev,\Martinec,\Ours}\ this means
the one-point functions as well as the partition
functions are completely determined by the integrable systems and symmetry
structures.
Our method can be generalized to the multi-point functions. Again, the
generating operators are determined by the recursion relations which comes
from the SD equations.

It is also possible to consider the $s_k$'s as variables such that one
can introduce another differential operators for the two-matrix model.
For the potentials $\V_1(U)=\sum_k t_k U^k, \V_2(V)=\sum_k s_k V^k$,
one can define
$$\eqalign{
&W^{(n+1)}_{m-n}Z\left[\{t_k\},\{s_k\}\right]\equiv -c^n\[\tr\(V^n U^m\)\],
\cr
&{\overline W}^{(n+1)}_{m-n}Z\left[\{t_k\},\{s_k\}\right]\equiv
-c^n\[\tr\(U^n V^m\)\],\cr}\eqn\eq$$
where both $W^{(n)}_{m}$ and ${\overline W}^{(n)}_{m}$ satisfy the recursion
relation Eq.\REQiv. The constraint equations are just
$$\left[c^m W^{(n+1)}_{m-n}- c^n {\overline W}^{(m+1)}_{n-m}\right]
Z\left[\{t_k\},\{s_k\}\right]=0.\eqn\eq$$

Another interesting point we want to mention is that the constraint
equations for the multi-matrix models like Eq.\MCEQ\ have very similar
form as those of the two-matrix models Eq.\CEQii. The coefficients of
the second potential of the two-matrix model can be decided by those of
the multi-matrix potentials. The correspondence, however, is
not quite exact. There appear some terms in the constraint equations of
the multi-matrix models which do not exist in those of the two-matrix
model. This observation reminds us of the recent conjectures that all
the multi-critical points can be achieved by the two-matrix
model.\refmark\Tada\ If the extra terms at finite $N$ are suppressed in the
double scaling limit, our observation can be a proof of this claim.
Related to this and other motivations, it would be very interesting to
consider the double scaling limit of our formalism.
Our discovery that the correlation functions are generated by the
$W_{1+\infty}$ algebra acting on the `$\tau$-functions' of
2D Toda hierarchy seems to be consistent with the results in the double
scaling limit in
that the correlation functions of the one-matrix model are
given by KdV-hierarchy equations\refmark\Rutgers\
and that the $\tau$-function of the $p$ reduced KP-hierarchy satisfy
the $W_{1+p}$ constraint equations.\refmark\Goeree\

Recently, there have been several papers which mention the
$W_{1+\infty}$ algebra. Our $W_{1+\infty}$ algebra is different from
that of ref.[\Itoyamaii] in that the latter comes from the higher order
terms under the change of $M$ to $M+\delta M$. Therefore, this
constraints exist even for the one-matrix model. Our $W_{1+\infty}$
constraints exist only for the multi-matrix models and will have direct
connection with the $W_n$ algebra structures conjectured in [\Kawai] in
the double scaling limit. Also, the $W_{1+\infty}$ algebra
appears from the KP hierarchy in the double scaling limit.\refmark\Sin\
This is a direct $p\to\infty$ limit of the $W_{1+p}$ constraint
equations considered in [\Goeree]. It would be interesting to consider the
$p\to\infty$ limit of our result to understand these results from the
matrix model point of view.

\noindent
{\bf Note}: While typing this paper, we received a paper [\Cheng]
from Y.-X. Cheng
where constraint equations for the two-matrix model like Eq.(30) have
been obtained.

\ack

C.A. thanks Tezukayama Univ. and Yukawa Institute of Kyoto Univ. at Uji
for their hospitality and financial support where part of this work has
been done.

{\singlespace
\refout}
\endpage
\end